 \documentclass[manuscript, screen]{acmart} 
 \usepackage{array}
 \usepackage{multirow}
 \usepackage{multicol}
 \usepackage{longtable}
 \usepackage{graphicx}
\pdfoutput = 1
\AtBeginDocument{%
  \providecommand\BibTeX{{%
    \normalfont B\kern-0.5em{\scshape i\kern-0.25em b}\kern-0.8em\TeX}}}

\setcopyright{acmcopyright}
\copyrightyear{2018}
\acmYear{2018}
\acmDOI{XXXXXXX.XXXXXXX}

\acmConference[Conference acronym 'XX]{Make sure to enter the correct
  conference title from your rights confirmation emai}{June 03--05,
  2018}{Woodstock, NY}
\acmPrice{15.00}
\acmISBN{978-1-4503-XXXX-X/18/06}

\begin{document}


\title{Predicting Opioid Use Outcomes in Minoritized Communities}



\author{Abhay Goyal}
\affiliation{%
  \institution{Dept. of CS, Missouri S \& T} 
  \country{USA}
}

\author{Nimay Parekh}
\affiliation{%
 \institution{Jivox, USA}
  \country{USA}}

\author{Lam	Yin Cheung}
\affiliation{%
 \institution{Yale University, USA} 
 \country{USA}}

 \author{Koustuv	Saha }
\affiliation{
 \institution{University of Illinois Urbana-Champaign, USA}
 \country{USA}}

 \author{Frederick L	Altice}
\affiliation{
 \institution{Yale University, USA}
 \country{USA}}

 \author{Robin O'hanlon}
\affiliation{%
 \institution{ Memorial Sloan Kettering Cancer Center, USA}
 \country{USA}}

\author{Roger Ho Chun Man}
\affiliation{%
 \institution{Institute for Health Innovation and Technology (iHealthtech), NUS, Singapore}
 \country{Singapore}}

\author{Christian Poellabauer}
\affiliation{%
 \institution{Florida International University, USA}
 \country{USA}}

\author{Honoria	Guarino}
\affiliation{%
 \institution{CUNY, USA}
 \country{USA}}

 \author{Pedro Mateu	Gelabert}
\affiliation{%
 \institution{CUNY, USA}
 \country{USA}}
 
 \author{ Navin Kumar}
\affiliation{%
 \institution{New York University, USA}
 \country{USA}}


\renewcommand{\shortauthors}{XX et al.}

\begin{abstract}
Machine learning algorithms can sometimes exacerbate health disparities based on ethnicity, gender, and other factors. There has been limited work at exploring potential biases within algorithms deployed on a small scale, and/or within minoritized communities. Understanding the nature of potential biases may improve the prediction of various health outcomes. As a case study, we used data from a sample of 539 young adults from minoritized communities who engaged in nonmedical use of prescription opioids and/or heroin. We addressed the indicated issues through the following contributions: 1) Using machine learning techniques, we predicted a range of opioid use outcomes for participants in our dataset; 2) We assessed if algorithms trained only on a majority sub-sample (e.g., Non-Hispanic/Latino, male), could accurately predict opioid use outcomes for a minoritized sub-sample (e.g., Latino, female). Results indicated that models trained on a random sample of our data could predict a range of opioid use outcomes with high precision. However, we noted a decrease in precision when we trained our models on data from a majority sub-sample, and tested these models on a minoritized sub-sample. We posit that a range of cultural factors and systemic forms of discrimination are not captured by data from majority sub-samples. Broadly, for predictions to be valid, models should be trained on data that includes adequate representation of the groups of people about whom predictions will be made. Stakeholders may utilize our findings to mitigate biases in models for predicting opioid use outcomes within minoritized communities.  
\end{abstract}




\maketitle

\section{Introduction}
Within the healthcare space, machine learning algorithms can sometimes exacerbate racial, ethnic, and gender disparities, among others. Many machine learning algorithms are trained on data from majority populations, thereby generating less accurate or reliable results for minoritized groups \cite{kostick2022mitigating}. For example, in a widely used algorithm, at a given risk score, the technique falsely concludes that Black individuals are healthier than equally sick White individuals \cite{obermeyer2019dissecting}. Thus, such large-scale algorithms can often perpetuate biases. There has been limited work at exploring potential biases in algorithms deployed within minoritized communities. In particular, minimal research has detailed how biases may manifest in algorithms developed by insurance companies to predict opioid use outcomes, or opioid overdoses among people who use opioids in urban areas. An algorithm trained on data from white individuals may provide incorrect estimates for Hispanic/Latino individuals, perhaps resulting in adverse health outcomes. 

Since predicting opioid use outcomes is important to improving health in populations often neglected by larger health systems \cite{kumar2022substance}, our goal is to examine how machine learning algorithms perform at determining opioid use outcomes within minoritized communities. As a case study, we used data from a sample of 539 young adults who engaged in nonmedical use of prescription opioids and/or heroin \cite{mateu2015injection}. The prevalence and incidence of opioid use has increased rapidly in the US in the past two decades, which is related to concomitant increases in opioid dependence, accidental overdose and death. We addressed the indicated issues through the following contributions: 1) Using machine learning techniques, we predicted opioid use outcomes for participants in our dataset; 2) We assessed if an algorithm trained on a majority sub-sample e.g., Non-Hispanic/Latino, male, could accurately predict opioid use outcomes for a minoritized sub-sample e.g., Latino, female. Our analysis was conducted to replicate possible real-world scenarios, and provide insight on how to improve broad health outcomes via predictive modeling. For example, if an insurance company primarily caters to Non-Hispanic/Latino individuals, models trained on data from Non-Hispanic/Latino individuals may not predict life insurance costing accurately for Hispanic individuals seeking treatment, and our analysis can provide understanding into such scenarios. 

\section{Related Work}
Several papers explored the creation of frameworks to mitigate biases within algorithms used in the healthcare space. A paper indicated that while technical AI advances in healthcare showcased impressive performances in lab settings, they seemed to fail when moving to clinical practice \cite{yildirim2021technical}. The authors discussed three interrelated challenges to AI in clinical practice: technical feasibility, financial viability, and clinician acceptance. Similar work explored queer concerns in privacy, censorship, language, online safety, and health \cite{tomasev2021fairness}. Most current approaches for algorithmic fairness assumed that the target characteristics for fairness, such as, race and legal gender, can be observed or recorded. The authors highlighted the importance of developing new approaches for algorithmic fairness that are different from the prevailing assumption of observed characteristics. Another paper indicated that AI-based applications raised not only ethical but legal and safety concerns \cite{kostick2022mitigating}, where models trained on data from majority sub-samples could generate less accurate or reliable results for minorities and other disadvantaged groups. Research has developed algorithms for a range of healthcare applications, such as to classify key types of gastrointestinal tract conditions \cite{thambawita2020extensive}, and to classify skin lesions \cite{lu2022fair}. Several studies have also used machine learning techniques to predict substance use outcomes. One study evaluated prediction performance of three different machine learning techniques in predicting opioid misuse among US adolescents, using the 2015–2017 National Survey on Drug Use and Health (NSDUH) \cite{han2020using}. Prediction models were developed using three algorithms, including neural networks, distributed random forest, and gradient Boosting. Findings suggested that machine learning techniques could be useful for predicting adolescent opioid misuse. Another study used a random forest classifier and NSDUH data to predict adults at risk for opioid use disorder (OUD) \cite{wadekar2019predicting}. Early initiation of cannabis, prior to 18 years of age, emerged as the dominant predictor for developing OUD in adult life. Other work used administrative claims data of 560,057 Medicare beneficiaries \cite{lo2019evaluation}, and found that deep neural networks and gradient boosting machine models outperformed other methods for identifying opioid overdose risk. 

Overall, research around whether machine learning exacerbates disparities within healthcare was limited, especially among smaller-scale cases. Most studies centered on developing frameworks to mitigate biases, or explored machine learning algorithms for various healthcare use cases with limited discussion on possible biases that may arise. We thus predicted a range of opioid use outcomes for minoritized sub-samples, and more importantly, determined how biases may occur in our use case.  

\section{Data}

\newcolumntype{L}{>{\arraybackslash}m{5cm}}
\begin{table}[htbp]
    \centering
    \begin{tabular}{|L|L|}
    \hline
    Sociodemographic Feature & Breakdown \\ \hline
    \multirow{2}{*}{Gender} &  Male (N=365) \\ \cline{2-2}
     & Female (N=170)  \\ \hline
    \multirow{2}{*}{Ethnicity} & Non-Hispanic/Latino (N=383) \\ \cline{2-2}
    & Hispanic/Latino (N=154)  \\ \hline
    \multirow{3}{*}{Race}  
    & White (N=497) \\ \cline{2-2}
    & Asian (N=7)  \\ \cline{2-2}
    & Black/African American (N=42)\\ \hline
    \multirow{2}{*}{Housing status}  & Not homeless (N=400) \\ \cline{2-2}
    & Homeless (N=137)  \\ \hline
    \multirow{2}{*}{Household income growing up} & Income$>$25000 (N=136) \\ \cline{2-2}
    & Income$<$25000 (N=390)  \\ \hline

    \end{tabular}
    \caption{Sociodemographic breakdown. Rows do not always add to 539 as we displayed larger subcategories.}
    \label{tab:demographics}
\end{table}

539 participants were recruited from July 2014 - October 2015 using respondent driven sampling, a form of chain-referral sampling designed to engage hard-to-reach populations. Most participants (Table \ref{tab:demographics}) were lower income ($<$25000/year) Non-Hispanic white men (N=365) who were not homeless were over-represented in the data. Women (N=170), Hispanic/Latino (N=154), Asian (N=7), Black (N=42), homeless (N=137), and high income participants (N=136) were underrepresented. The questionnaire used the standard items for collecting race and ethnicity as per the 1997 Office of Management and Budget (OMB) Revisions to the Standards for the Classification of Federal Data on Race and Ethnicity. These items are commonly used for federal data collection purposes. We conformed to these race and ethnicity criteria as the data collection process was federally funded. The revised OMB standards contained five minimum categories for race: American Indian or Alaska Native, Asian, Black or African American, Native Hawaiian or Other Pacific Islander, and white. There were two categories for ethnicity: Hispanic or Latino and Not Hispanic or Latino. Thus, we collected both race and ethnicity data. Participants were asked to recruit fellow prescription opioid and/or heroin users. Several papers have been published with this dataset and further information on data collection and methods was contained in past work \cite{friedman2017sexual,mateu2015injection}. Eligibility criteria included: nonmedical use of prescription opioids and/or heroin use in the past 30 days; current residence in NYC; aged 18–29 years-old; English-speaking; and ability to provide informed consent. Structured interviews lasted between 90 and 120 minutes and were used to collect quantitative data on drug use, sexual behavior, sexual and injection partnerships and networks, overdose, drug treatment, HIV and hepatitis C knowledge and testing history. 

\section{Methods}
\textbf{Features of interest.}
\newcolumntype{L}{>{\arraybackslash}m{8cm}}
\begin{table}[h]
    \centering
    \begin{tabular}{|L|L|}
    
    \hline
    Outcome &  Possible values\\
    \hline



Have you injected drugs in the past 3 months? & No (N=222), Yes (N=317)  \\ \hline
       
Have you ever been in treatment for drug use? &  No (N=155), Yes (N=384)\\ 

              \hline

    \end{tabular}
    \caption{Opioid use outcomes of interest}
    \label{tab:information}
\end{table}

\newcolumntype{k}{>{\arraybackslash}m{4cm}}
\newcolumntype{L}{>{\arraybackslash}m{8cm}}
\begin{table}[htbp]
    \centering
    \begin{tabular}{|k|L|k|}
    \hline

    Category & Feature & Possible values \\ \hline 

     \multirow{20}{*}{Sociodemographics} & 
    1. Do you consider yourself to be Hispanic or Latino/a? & No, Yes \\ \cline{2-3}
    & \multirow{4}{*}{2. Which racial group do you consider yourself to be in?}
    &  a. American Indian \\ \cline{3-3}
    && b. Asian \\ \cline{3-3}
    && c. Black/African American\\ \cline{3-3}
    && d. Native Hawaiian/Pacific Islander \\ \cline{3-3}
    && e. White 
    \\ \cline{2-3}
    & \multirow{5}{*}{\parbox{8cm}{3. What borough do you currently live in?}} 
    &  a. Manhattan \\ \cline{3-3}
    && b. Staten Island\\ \cline{3-3}
    && c. Brooklyn\\ \cline{3-3}
    && d. Bronx\\ \cline{3-3}
    && e. Queens \\ \cline{2-3}
    
    & \multirow{5}{*}{\parbox{8cm}{4. Do you consider yourself to be male, female or transgender?}} 
    &  a. Male \\ \cline{3-3}
    && b. Female\\ \cline{3-3}
    && c. Transgender Male to Female\\ \cline{3-3}
    && d. Transgender Female to Male\\  \cline{2-3}
    
 \hline
     \multirow{6}{*}{\parbox{4cm}{ Sexual Violence}} 
    
    &  1. Have you ever had the experience while you were using drugs when someone touched or rubbed against your body in a sexual manner without your consent? &  No, Yes \\ \cline{2-3}

    & 2. Have you ever felt that someone expected you to have sex with them because you were using drugs together? & No, Yes  \\ \cline{2-3}

    &  {\parbox{8cm}{ 3. While you were using drugs, did someone ever have sex (oral,
vaginal, and/or anal) with you without your consent?}} & No, Yes \\ \cline{2-3}

    
  \hline

    \multirow{15}{*}{\parbox{4cm}{ Perceived Risk for HIV, HCV \& STIs}} 
    
      &   \multirow{5}{*}{1. How concerned are you about getting HIV/AIDS? }
        &a. Not At All Concerned \\ \cline{3-3}
        && b. Somewhat Concerned \\ \cline{3-3}
        && c .Very Concerned \\ \cline{3-3}
        && d. I Have It\\ \cline{2-3}
       & \multirow{4}{*}{2. How concerned are you about getting Hepatitis C? } 
       &   a. Not At All Concerned \\ \cline{3-3}
            && b. Somewhat Concerned \\ \cline{3-3}
            && c. Very Concerned \\ \cline{3-3}
            && d. I have it \\ \cline{2-3}
        &  \multirow{4}{*}{3. How concerned are you about getting chlamydia?} 
            & a.Not At All Concerned \\ \cline{3-3}
            && b. Somewhat Concerned\\ \cline{3-3}
            && c. Very Concerned\\ \cline{3-3}
            && d. I have it\\ \cline{2-3}
             \hline

    \end{tabular}


    \centering
    \begin{tabular}{|k|L|k|}

      \multirow{5}{*}{ Mental Health}
    & 1. Have you ever been diagnosed with a psychological or
emotional problem? & No, Yes \\ \cline{2-3}
        & 2. How many times in your life, if any, have you been treated for
any psychological or emotional problem in a hospital or inpatient
unit?  & No. Of Episodes \\ \cline{2-3}
        & 3. Experienced serious depression? & No, Yes \\ 
  \hline

    \end{tabular}
    \caption{Features of interest}
    \label{tab:features}
\end{table}


\textit{Opioid use outcomes of interest.}
To develop a list of outcomes to predict, we first selected content experts with a research specialization in opioid use who assembled a list of features relevant to opioid use outcomes. Content experts had published >10 peer-reviewed articles in opioid use, broadly defined. All outcomes were derived from the quantitative survey data collected. Content experts only retained items that had been agreed upon after discussion with any disagreements resolved by a third content expert. Our final list of two outcomes that we sought to predict using machine learning are in Table \ref{tab:information}.
\textit{Feature selection.}
Content experts selected an initial list of questions for feature selection. All features were derived from the survey questions. Content experts only retained items that had been agreed upon after discussion. Experts only selected items that did not replicate the opioid use outcomes of interest. For example, the question \textit{In the past 30 days, did you take prescription opioid(s) nonmedically?} was not included in our initial set of features as it was too similar to the opioid use outcomes we were trying to predict. Any disagreements were resolved by a third content expert. This process resulted in 327 features. Following this, we used the following feature selection techniques to reduce the feature set: Coefficient of variation (CV); Pairwise correlations (PC). CV: We employed the CV method to reduce the feature space based on explained variance. This involved measuring the ratio of standard deviation to mean for each feature and removing any outliers that exceeded a threshold CV of two standard deviations from the mean (21.67). Two features exceeded this threshold and were subsequently dropped. PC: Correlated features which can negatively impact machine learning prediction models by producing unstable solutions or obscuring the interactions between significant features. To identify uncorrelated features, we calculated Pearson's correlation coefficient between all feature pairs and removed any with an absolute correlation value above the threshold of 0.8. This resulted in the removal of 74 features. Examples from the final list of 280 selected features were presented in Table \ref{tab:features}. Feature names and options were verbatim from the interviews. Missing data was imputed with a value of 0.

\textbf{Models.}
We selected several commonly used models to predict opioid use outcomes, as below. \\
\textit{Random Forest Classifier.}
Random forest classifier is an ensemble machine learning method that is computationally efficient over large datasets. The Random forest uses multiple decision trees and then finds the most predicted class when making the prediction. We used an estimator size of 100 and depth to 100 to allow for pruning to take place. For initial feature prediction, we used a train size of 70\% and test size of 30\%.\\
\textit{Bagging Classifier.}
Bagging classifier is an ensemble method which uses other basic methods, in our case, decision trees and \textit{bags} or combines the different results together. It then chooses the predicted label as the one which has been predicted by most of the decision trees. In bagging classifier, we used a 70\% and 30\% train/test split.\\ 
\textit{Gradient Boosting Classifier.}
The gradient boosting classifier builds different models sequentially and it builds the next model such that the losses of the previous one are minimized. The loss used in the classifier is the log-likelihood.\\
\textit{Adaptive Boosting Classifier.}
The adaptive boosting classifier (AdaBoost) is made such that it adapts based on errors made by the previous classifier. The output of the other learning algorithms is combined into a weighted sum that represents the final output of the boosted classifier. We used the decision trees as the base estimator with a depth of 100.\\

\textbf{Feature prediction.}
\newcolumntype{L}{>{\arraybackslash}m{5cm}}
\begin{table}[htbp]
    \centering
    \begin{tabular}{|L|L|L|}
    \hline
    Feature  & Majority Sociodemographic Group & Minority Sociodemographic Group \\ \hline
    
     Do you consider yourself to be male, female or transgender? & Male (N=365) &  Female (N=170)  \\ \hline
    
    Do you consider yourself to be Hispanic or Latino/a & Non-Hispanic/Latino (N=383) & Hispanic/Latino (N=154)  \\ \hline
    \end{tabular}
    \caption{Features to subset data}
    \label{tab:information1}
\end{table}

We first predicted each of the outcomes in Table \ref{tab:information} using all 280 features, some of which were indicated in Table \ref{tab:features}. We kept to guidelines which indicated \textit{n-1} as the optimal feature size for uncorrelated features \cite{hua2005optimal}. We then predicted the same outcomes, but trained our models on a majority sub-sample, then tested the models on a individual minoritized sub-sample, based on the features in Table \ref{tab:information1}. Content experts selected majority and minoritized/minority sub-samples based on where the greatest discrepancy in opioid use outcomes may exist between two groups, e.g., male/female. By majority and minoritized/minority, we referred to these demographic groups' representation within this particular sample, not necessarily their representation in society at large. We mainly referred to the distribution of these groups within this sample, but used the data to make a larger point about the use of machine learning algorithms to predict health risk in a broad range of US populations. A group's status as \textit{majority} versus \textit{minority} was context-dependent; for example, while females were a minority in this sample, they were not a minority in the general US population. For each outcome that we were trying to predict, we trained our model on only data from a single sub-sample indicated in the \textit{Majority Sociodemographic Group} column of Table \ref{tab:information1}, and tested the model on data from a single sub-sample detailed in the \textit{Minority Sociodemographic Group} column of Table \ref{tab:information1}. We repeated this process with each minoritized/minority sub-sample pair in Table \ref{tab:information1}, and outcome as per Table \ref{tab:information}. For example, in predicting the variable \textit{Have you injected drugs in the past 3 months?}, we trained our models only on data from Non-Hispanic/Latino participants (N=383), but tested models with data only from Hispanic/Latino participants (N=154). We then did the same, but with male (N=365) and female participants (N=170). 

\section{Results}

\newcolumntype{L}{>{\arraybackslash}m{5cm}}
\begin{table}[htbp]
    \hspace*{-1cm}
    \centering
    \small
    \label{tab:results1}
    \begin{tabular}{|p{5cm}|c|c|c|c|c|}
        \hline
        \multirow{2}{*}{\parbox{5cm}{\centering Feature}} & \multirow{2}{*}{Model} & \multicolumn{4}{c|}{Metrics} \\ \cline{3-6}
         & & ROC AUC & Accuracy & F1-Score & Average Precision \\ \hline
         
    \multirow{4}{*}{\parbox{5cm}{ Have you injected drugs in the past 3 months?}} & Random Forest & 0.785 & 0.802 & 0.802 & 0.775 \\ \cline{2-6}
     & Adaptive Boost & 0.809 & 0.820 & 0.820 & 0.799 \\ \cline{2-6}
     & Gradient Boost & 0.825 & 0.839 & 0.839 & \textbf{0.811} \\ \cline{2-6}
     & Bagging classifier & 0.757 & 0.759 & 0.759 & 0.759 \\ \cline{2-6} \hline

     \multirow{4}{*}{\parbox{5cm}{ Have you ever been in treatment for drug use?}} & Random Forest & 0.695 & 0.777 & 0.777 & 0.787 \\ \cline{2-6}
     & Adaptive Boost & 0.710 & 0.753 & 0.753 & 0.798 \\ \cline{2-6}
     & Gradient Boost & 0.747 & 0.796 & 0.796 &\textbf{ 0.819} \\ \cline{2-6}
     & Bagging classifier & 0.734 & 0.777 & 0.777 & 0.811 \\ \cline{2-6} \hline


    \end{tabular}
       
        \caption{Predictions for opioid use outcomes}
         \label{tab:results1}
\end{table}

We first used a range of models to predict opioid use outcomes as per Table \ref{tab:results1}. While we provided several model metrics, we believed that average precision provided the most insight as precision is the best metric to be used when the classes are highly imbalanced \cite{armah2014deep}. Precision is the fraction of instances that the model classifies as positive that are indeed correct. We indicated in \textbf{bold} the highest precision for each feature we predicted. Moreover, precision provided the proportion of positive identifications which were accurately predicted, central to our analysis. Based on all metrics for each model per feature, we indicated in \textbf{bold} the best performing model for each feature. Overall, our models predicted the opioid use outcomes of interest with relatively high precision, likely due to the survey questions being specifically designed to understand opioid use outcomes and the fact that the survey was administered to a substance-using population.

\begin{figure*}
    \centering
    (a) Injection drug use in the last three months \label{fig:Injection drug use in the last three months}
    {{\includegraphics[height=8cm, width=15cm]{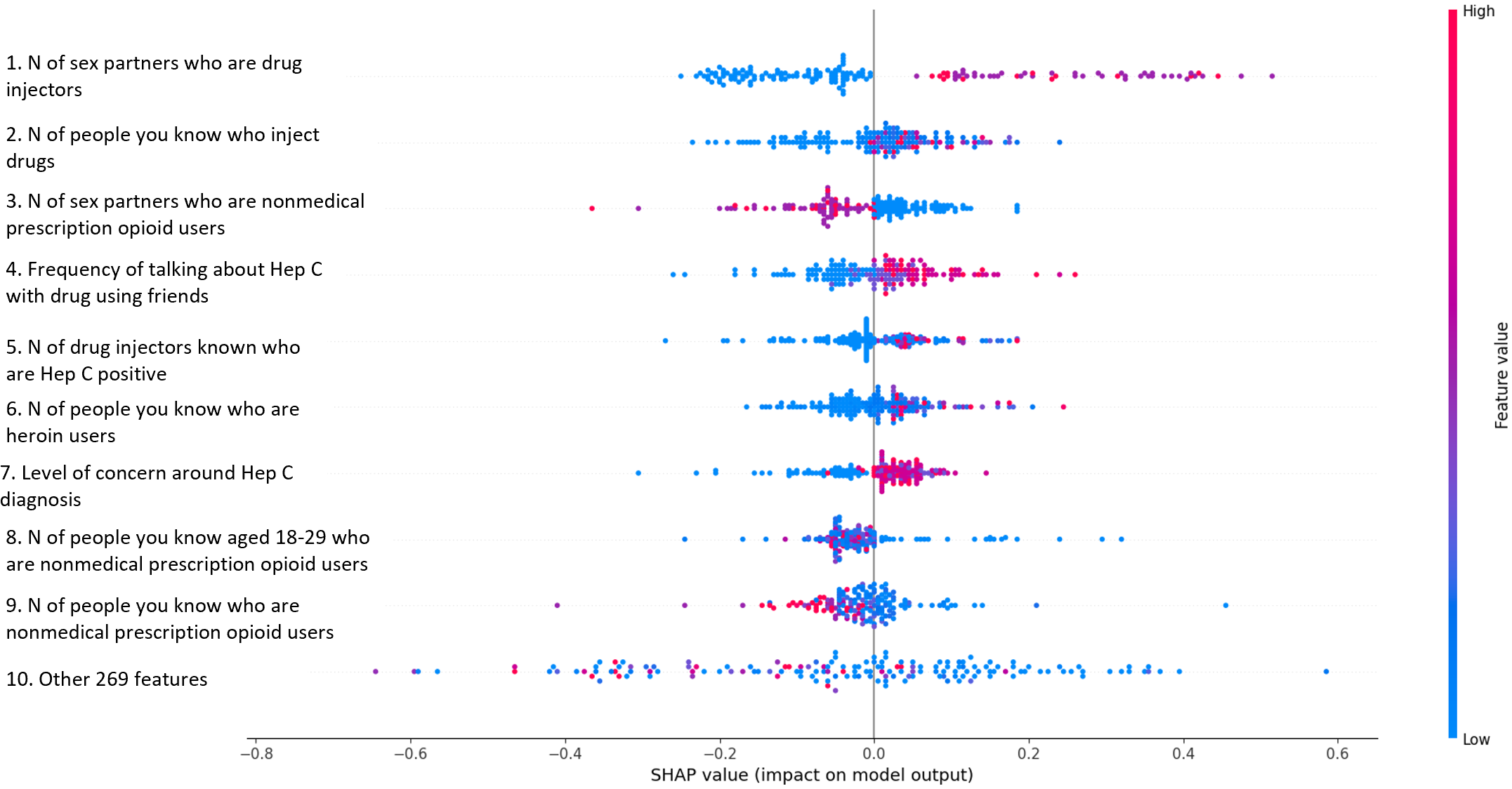}}}\qquad
    \hspace*{-0.5cm}
    (b) Being in treatment for drug use \label{fig:Being in treatment for drug use}
    {{\includegraphics[height=8cm, width=15cm]{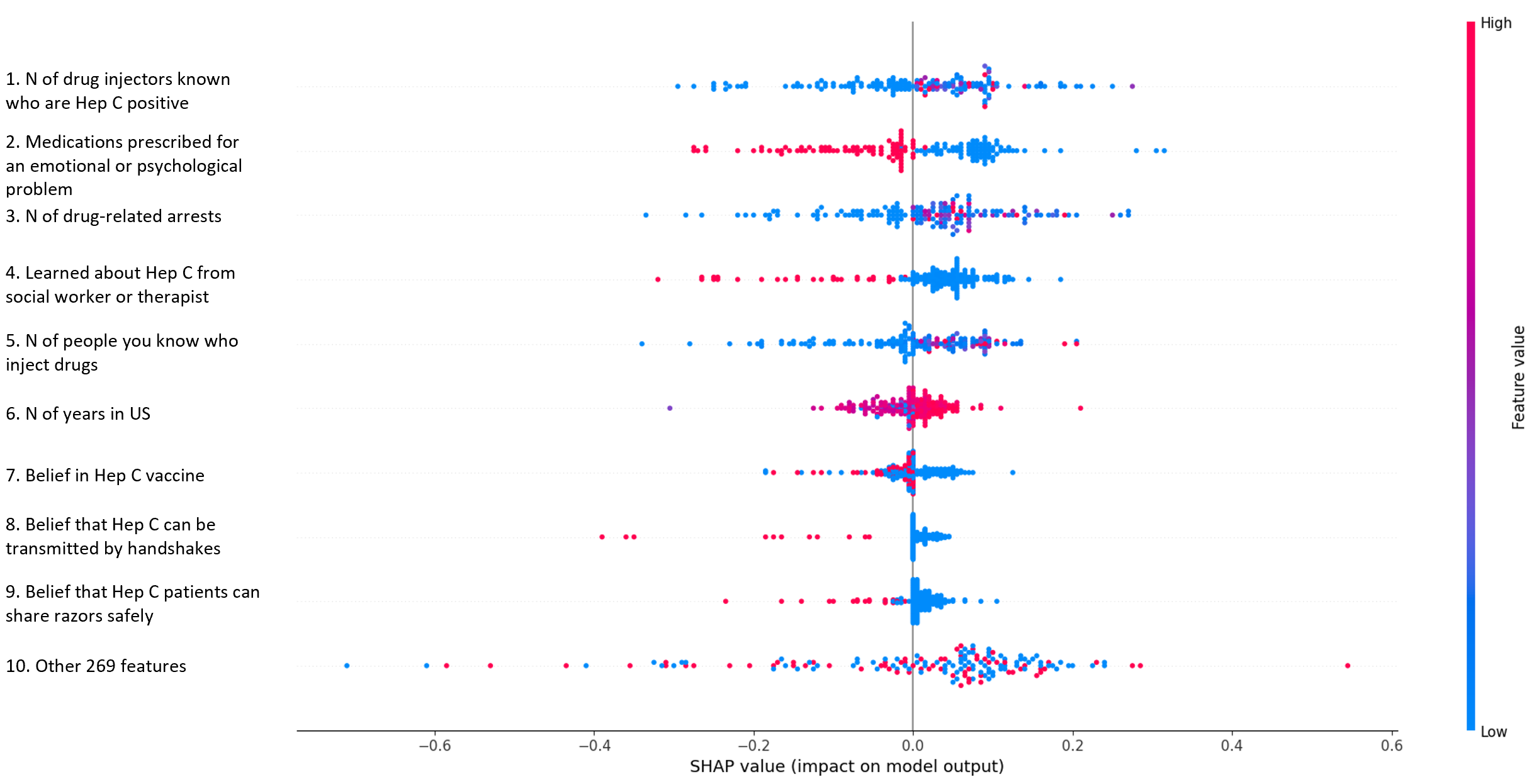}}}\qquad
\caption{SHAP visualizations for opioid use outcomes} \label{fig:beeswarm}
\end{figure*}

To provide more insight into our results, we used SHapley Additive exPlanations (SHAP) to understand how each feature contributed to predicting the opioid use outcomes. SHAP quantifies the contribution each feature brings to the prediction made by the model. We detailed beeswarm plots (Figure \ref{fig:beeswarm}) for the models with the highest precision, for each opioid use outcome predicted. Features that contributed the most to the prediction are closer to the top of the plot. Broadly, having peers who engaged in opioid use seemed to contribute most to the prediction of being in treatment for drug use, and injecting drugs. Past work suggested the role of peer effects in initiating opioid use \cite{henneberger2021peer}. Peer effects are when one's behavior is affected by the behavior of their peers. However, apart from peer effects, there were likely other features which more accurately predict opioid use. For example, engagement in harm reduction practices, such as using clean syringes, or proximity to a methadone treatment center, may relate to injecting drugs, but such features were not collected in the data. 

\begin{table}[h]
    \small
    \begin{tabular}{|p{3cm}|l|l|l|l|l|l|}
    \hline

    \multirow{3}{*}{\parbox{3cm}{ Feature}} & \multirow{3}{*}{Majority}  & \multirow{3}{*}{Minority} 
        &\multirow{3}{*}{Model} &\multicolumn{3}{c|}{Metrics} \\ 
        \cline{5-7} 
         & & & & Accuracy & F1-Score & Average Precision \\ \hline
         
    \multirow{13}{*}{\parbox{3cm}{ Have you ever injected drugs in the past 3 months?}}  & \multirow{8}{*}{\parbox{1cm}{ Male (N=369)}} & \multirow{8}{*}{\parbox{1cm}{ Female (N=170)}} 
    & Random Forest & 0.758 & 0.758 & 0.721 \\ \cline{4-7}
    ~ & ~ &  & Adaptive Boosting & 0.788 & 0.788 &\textbf{0.754} \\ \cline{4-7}
    ~ & ~ &  & Gradient Boosting & 0.794 & 0.794 & 0.750 \\ \cline{4-7}
    ~ & ~ &  & Bagging classifier & 0.723 & 0.723 & 0.728 \\
    \cline{4-7}
    \cline{2-7}

    &  \multirow{8}{*}{\parbox{1cm}{  Non-Hispanic/
    Latino (N=385)}} & \multirow{8}{*}{\parbox{1cm}{ Hispanic/
    Latino (N=154)}}  
    & Random Forest & 0.792 & 0.792 & 0.629 \\
    \cline{4-7}

    ~ & ~ &  & Adaptive Boosting & 0.831 & 0.831 & \textbf{0.674}\\ \cline{4-7}
    ~ & ~ &  & Gradient Boosting  & 0.805 & 0.805 & 0.639 \\ \cline{4-7}
    ~ & ~ &  & Bagging classifier & 0.798 & 0.798 & 0.628 \\ \hline

    \multirow{13}{*}{\parbox{3cm}{ Have you ever been in treatment for drug use?}}  & \multirow{8}{*}{\parbox{1cm}{ Male (N=369)}} & \multirow{8}{*}{\parbox{1cm}{ Female (N=170)}} 
    & Random Forest & 0.788 & 0.788 & 0.797 \\ \cline{4-7}
    ~ & ~ &  & Adaptive Boosting & 0.782 & 0.782 & 0.812 \\ \cline{4-7}
    ~ & ~ &  & Gradient Boosting & 0.709 & 0.8 & 0.8 \\ \cline{4-7}
    ~ & ~ &  & Bagging classifier & 0.764 & 0.764 & \textbf{0.816} \\
    \cline{4-7}
    \cline{2-7}

    &  \multirow{8}{*}{\parbox{1cm}{  Non-Hispanic/
    Latino (N=385)}} & \multirow{8}{*}{\parbox{1cm}{ Hispanic/
    Latino (N=154)}}  
    & Random Forest & 0.720 & 0.720 & 0.673 \\
    \cline{4-7}

    ~ & ~ &  & Adaptive Boosting & 0.714 & 0.714 & 0.686\\ \cline{4-7}
    ~ & ~ &  & Gradient Boosting  & 0.727 & 0.727 & \textbf{0.688} \\ \cline{4-7}
    ~ & ~ &  & Bagging classifier & 0.694 & 0.694 & 0.667 \\ \hline

    \hline
\end{tabular}

\caption{Predictions for opioid use outcomes by training model on data from a majority sub-sample and testing model on data from a minority sub-sample}
\label{tab:results2}
\end{table}

We then sought to determine if models trained on a majority sub-sample could predict outcomes for a minoritized sub-sample (Table \ref{tab:results2}). We indicated in \textbf{bold} the highest precision for each feature we predicted. As expected, training models on a majority sub-sample to predict opioid use outcomes for a minoritized sub-sample provided poorer results than compared to our analysis in Table \ref{tab:results1}. We compared precision between our two sets of analysis (Table \ref{tab:results1}, Table \ref{tab:results2}). Essentially, we compared the differences in precision between: 1) training a model on a randomly selected subset of our data and testing the same model on a randomly selected subset of our data; 2) training a model on data from a majority sub-sample (e.g., Non-Hispanic/Latino participants), and testing the same model on data from a minoritized sub-sample (e.g., Hispanic/Latino participants). 

We noted a decrease in precision when predicting opioid use outcomes across the minoritized categories of interest. Results may suggest that using data from males to predict female opioid use, and predicting Hispanic/Latino opioid use with data from non-Hispanics/Latinos was infeasible. The reduced precision in predicting opioid use outcomes may be attributed to the specific patterns of drug use among young people in our dataset. Most of the participants in our study began using opioids during their teenage years, with prescription opioids (POs) being the primary form of nonmedical use. However, by the time they were included in the study (with an average age of 24.5), many had switched to using heroin instead of POs \cite{guarino2018young}. This trajectory was influenced by their birth cohort and the availability of different drugs during their adolescence. During the time when our sample were 16-17 years old, PO prescribing in the US was at an all-time high, and diverted POs were easily accessible. As POs became more challenging and expensive to obtain, those who developed opioid use disorder usually transitioned to using heroin, which was cheaper and more potent in New York City. There were also racial/ethnic differences in these patterns, with white individuals more likely to use heroin and have longer opioid use histories, while other groups were more likely to use POs only and have shorter and less severe opioid use histories. PO use was a recent phenomenon in the peer networks of young Black and Latino individuals at the time of data collection, but white communities had been affected by the opioid epidemic earlier, resulting in more prolonged exposure to opioids among white adolescents \cite{guarino2018young}. Furthermore, individuals who had longer histories of opioid use and more severe opioid use disorder, predominantly male and white individuals, were more likely to engage in the challenging process of breaking down PO pills for injection \cite{guarino2018young}. These differing behaviors across gender and ethnic/racial categories may elucidate why models trained on data from males did not accurately predict female injection use. Harm reduction policy approaches are generally lacking in low-income ethnic minority neighborhoods. Thus, datasets to predict opioid use may not have features which model harm reduction initiatives in such communities, perhaps resulting in poor precision to predict opioid use outcomes for Hispanic/Latino populations with models trained on a majority population. Similarly, unstable housing is a predictor for opioid use outcomes, but datasets may not contain data from unhoused people, or features that aggregate the experience of being unhoused. 

We suggest that researchers designing such models in diverse populations need to be aware of cultural nuances, such as differential opioid use behaviors across racial/ethnic groups. For example, training a model on data from one racial/ethnic community and using it to predict injection drug use in another community may perpetuate biases, and completely ignore racial/ethnic nuances around opioid use. While there have been numerous attempts to predict opioid use outcomes using a range of machine learning techniques \cite{tseregounis2021assessing}, models have a high rate of false positives, largely because opioid use outcomes such as overdose are rare in the general population, and some important, non-person-level factors contribute strongly to overdose - such as potency of a particular batch or dose of drugs. Moreover, these models are rarely implemented in clinical practice. Our results suggest that predicting opioid use outcomes within minoritized communities may have low precision, indicating that much more work is required before machine learning algorithms to predict such outcomes can be productionized in healthcare systems.

\section{Discussion}
\textbf{Implications of findings.} 
Our goal was to predict a range of opioid use outcomes as assessed in a sample of young people who use opioids, and determine if an algorithm trained only on a majority sub-sample could accurately predict opioid use outcomes for a minoritized sub-sample, to inform best practices for the use of predictive algorithms in substance-using populations. Models were able to predict recent injection drug use and participation in drug treatment. The presence of peers who also engaged in opioid use appeared to play a role in predicting drug treatment and injection drug use. However, the available data lacked comprehensive information on other facets of opioid use, such as harm reduction. We noted a decrease in precision when we trained our models on only data from a majority sub-sample, and tested these models on a minoritized sub-sample. Overall, machine learning approaches are only as precise and useful as the data they are trained on, and to make valid and accurate predictions they must be trained on data from people who are similar in terms of key sociodemographic characteristics as the populations about whom predictions will be made.\\
\textbf{Recommendations.}
Key to mitigating biases in models to predict health outcomes within minoritized communities, is the inclusion of stakeholders at every stage of the machine learning operations (MLOps) pipeline. For example, methadone patients need to be involved in the development of models to predict methadone dropout risk \cite{chen2022partisan,chen2023categorizing}. Similarly, a committee of ethnic minority individuals can be involved in auditing algorithms used to detect cardiovascular risk. Insurance companies and other stakeholders who use machine learning to predict opioid use outcomes need to be aware that models can exacerbate biases, and seek to improve their predictive modelling capabilities. Insurance companies that have primarily white individuals in their datasets should seek to augment their datasets with individuals from minoritized backgrounds. Such practices can aid providers in making accurate predictions if their client demographics shift, or if nonwhite individuals seek treatment. There increasingly exist independent corporations that audit large scale machine learning models, and such corporations need to ensure that minoritized communities are adequately represented within the audit committee. Similarly, governments and similar stakeholders need to initiate guardrails and implement sound AI-centric policymaking. Interventions to educate the public on such models is also critical. For example, stakeholders can develop short videos on YouTube explaining how such algorithms may reproduce biases. Including features in models that capture the nuances of opioid use behaviors within minoritized communities is key  \cite{oles2022characterizing,kumar2022online}. Models can include features that capture individuals' distance from harm reduction centers, or their engagement with harm reduction practices, such as the use of sterile syringes, and fentanyl testing. 

\textbf{Limitations.}
Findings relied on the validity of questions collected within the dataset, and there may have been features of interest not included in the questionnaire. Our data may not be generalizable to other health outcomes or populations. Questionnaire answers may be unreliable, perhaps further limiting generalizability. Our dataset is small relative to other similar studies, and we will collect a larger sample in future work. We were unable to conduct balancing for our data given the small sample size. The data pertains to a highly particular subset of people who use drugs, specifically, young adults residing in NYC who engage in opioid use. Therefore, it remains uncertain to what degree the outcomes can be applicable to individuals in dissimilar geographic regions, varying age groups, or those consuming different types of substances. 



\bibliographystyle{acm}

\bibliography{sample-base}

\begin{thebibliography}{10}

\bibitem{armah2014deep}
{\sc Armah, G.~K., Luo, G., and Qin, K.}
\newblock A deep analysis of the precision formula for imbalanced class
  distribution.
\newblock {\em International Journal of Machine Learning and Computing 4}, 5
  (2014), 417--422.

\bibitem{chen2022partisan}
{\sc Chen, K., Babaeianjelodar, M., Shi, Y., Janmohamed, K., Sarkar, R., Weber,
  I., Davidson, T., De~Choudhury, M., Yadav, S., Khudabukhsh, A., et~al.}
\newblock Partisan us news media representations of syrian refugees.
\newblock {\em arXiv preprint arXiv:2206.09024\/} (2022).

\bibitem{chen2023categorizing}
{\sc Chen, K., Feng, A., Aanegola, R., Saha, K., Wong, A., Schwitzky, Z., Lee,
  R. K.-W., O’Hanlon, R., De~Choudhury, M., Altice, F.~L., et~al.}
\newblock Categorizing memes about the ukraine conflict.
\newblock In {\em Computational Data and Social Networks: 11th International
  Conference, CSoNet 2022, Virtual Event, December 5--7, 2022, Proceedings\/}
  (2023), Springer, pp.~27--38.

\bibitem{friedman2017sexual}
{\sc Friedman, S., Mateu-Gelabert, P., Ruggles, K., Goodbody, E., Syckes, C.,
  Jessell, L., Teubl, J., and Guarino, H.}
\newblock Sexual risk and transmission behaviors, partnerships and settings
  among young adult nonmedical opioid users in new york city.
\newblock {\em AIDS and Behavior 21\/} (2017), 994--1003.

\bibitem{guarino2018young}
{\sc Guarino, H., Mateu-Gelabert, P., Teubl, J., and Goodbody, E.}
\newblock Young adults' opioid use trajectories: From nonmedical prescription
  opioid use to heroin, drug injection, drug treatment and overdose.
\newblock {\em Addictive behaviors 86\/} (2018), 118--123.

\bibitem{han2020using}
{\sc Han, D.-H., Lee, S., and Seo, D.-C.}
\newblock Using machine learning to predict opioid misuse among us adolescents.
\newblock {\em Preventive medicine 130\/} (2020), 105886.

\bibitem{henneberger2021peer}
{\sc Henneberger, A.~K., Mushonga, D.~R., and Preston, A.~M.}
\newblock Peer influence and adolescent substance use: A systematic review of
  dynamic social network research.
\newblock {\em Adolescent Research Review 6\/} (2021), 57--73.

\bibitem{hua2005optimal}
{\sc Hua, J., Xiong, Z., Lowey, J., Suh, E., and Dougherty, E.~R.}
\newblock Optimal number of features as a function of sample size for various
  classification rules.
\newblock {\em Bioinformatics 21}, 8 (2005), 1509--1515.

\bibitem{kostick2022mitigating}
{\sc Kostick-Quenet, K.~M., Cohen, I.~G., Gerke, S., Lo, B., Antaki, J.,
  Movahedi, F., Njah, H., Schoen, L., Estep, J.~E., and Blumenthal-Barby, J.}
\newblock Mitigating racial bias in machine learning.
\newblock {\em Journal of Law, Medicine \& Ethics 50}, 1 (2022), 92--100.

\bibitem{kumar2022online}
{\sc Kumar, N., Chen, K., Shi, Y., and Altice, F.~L.}
\newblock Online platforms' framing around vaping.
\newblock {\em Drug Testing and Analysis\/} (2022).

\bibitem{kumar2022substance}
{\sc Kumar, N., Janmohamed, K., Nyhan, K., Martins, S.~S., Cerda, M., Hasin,
  D., Scott, J., Frimpong, A.~S., Pates, R., Ghandour, L.~A., et~al.}
\newblock Substance, use in relation to covid-19: a scoping review.
\newblock {\em Addictive behaviors 127\/} (2022), 107213.

\bibitem{lo2019evaluation}
{\sc Lo-Ciganic, W.-H., Huang, J.~L., Zhang, H.~H., Weiss, J.~C., Wu, Y., Kwoh,
  C.~K., Donohue, J.~M., Cochran, G., Gordon, A.~J., Malone, D.~C., et~al.}
\newblock Evaluation of machine-learning algorithms for predicting opioid
  overdose risk among medicare beneficiaries with opioid prescriptions.
\newblock {\em JAMA network open 2}, 3 (2019), e190968--e190968.

\bibitem{lu2022fair}
{\sc Lu, C., Lemay, A., Chang, K., H{\"o}bel, K., and Kalpathy-Cramer, J.}
\newblock Fair conformal predictors for applications in medical imaging.
\newblock In {\em Proceedings of the AAAI Conference on Artificial
  Intelligence\/} (2022), vol.~36, pp.~12008--12016.

\bibitem{mateu2015injection}
{\sc Mateu-Gelabert, P., Guarino, H., Jessell, L., and Teper, A.}
\newblock Injection and sexual hiv/hcv risk behaviors associated with
  nonmedical use of prescription opioids among young adults in new york city.
\newblock {\em Journal of substance abuse treatment 48}, 1 (2015), 13--20.

\bibitem{obermeyer2019dissecting}
{\sc Obermeyer, Z., Powers, B., Vogeli, C., and Mullainathan, S.}
\newblock Dissecting racial bias in an algorithm used to manage the health of
  populations.
\newblock {\em Science 366}, 6464 (2019), 447--453.

\bibitem{oles2022characterizing}
{\sc Oles, W., Alexander, M., Kumar, N., Howell, B., O’Connor, P.~G., Madden,
  L.~M., and Barry, D.~T.}
\newblock Characterizing the social support and functioning of a low-threshold
  medication for opioid use disorder treatment cohort at intake.
\newblock {\em BMC psychiatry 22}, 1 (2022), 236.

\bibitem{thambawita2020extensive}
{\sc Thambawita, V., Jha, D., Hammer, H.~L., Johansen, H.~D., Johansen, D.,
  Halvorsen, P., and Riegler, M.~A.}
\newblock An extensive study on cross-dataset bias and evaluation metrics
  interpretation for machine learning applied to gastrointestinal tract
  abnormality classification.
\newblock {\em ACM Transactions on Computing for Healthcare 1}, 3 (2020),
  1--29.

\bibitem{tomasev2021fairness}
{\sc Tomasev, N., McKee, K.~R., Kay, J., and Mohamed, S.}
\newblock Fairness for unobserved characteristics: Insights from technological
  impacts on queer communities.
\newblock In {\em Proceedings of the 2021 AAAI/ACM Conference on AI, Ethics,
  and Society\/} (2021), pp.~254--265.

\bibitem{tseregounis2021assessing}
{\sc Tseregounis, I.~E., and Henry, S.~G.}
\newblock Assessing opioid overdose risk: a review of clinical prediction
  models utilizing patient-level data.
\newblock {\em Translational Research 234\/} (2021), 74--87.

\bibitem{wadekar2019predicting}
{\sc Wadekar, A.}
\newblock Predicting opioid use disorder (oud) using a random forest.
\newblock In {\em 2019 IEEE 43rd Annual Computer Software and Applications
  Conference (COMPSAC)\/} (2019), vol.~1, IEEE, pp.~960--961.

\bibitem{yildirim2021technical}
{\sc Yildirim, N., Zimmerman, J., and Preum, S.}
\newblock Technical feasibility, financial viability, and clinician acceptance:
  On the many challenges to ai in clinical practice.
\newblock In {\em HUMAN@ AAAI Fall Symposium\/} (2021).

\end{thebibliography}


\end{document}